\documentstyle[amssymb,pra,aps,epsf]{revtex}

\begin{document}
\title{Singular potentials and annihilation.}
\author{A. Yu. Voronin}
\address{P. N. Lebedev Physical Institute\\
53 Leninsky pr.,117924 Moscow, Russia}
\maketitle

\begin{abstract}
We discuss the relation between attractive singular potentials $-\alpha
_{s}/r^{s}$, $s\geq 2$ and inelasticity. We show that mentioned singular
potentials can be regularized by infinitesimal imaginary addition to
interaction constant $\alpha _{s}=\alpha _{s}\pm i0$. Such a procedure
enables unique definition of scattering observables and is equal to an
absorption (creation) of particles in the origin. It is shown, that
suggested regularization is an analytical continuation of the scattering
amplitudes of repulsive singular potential as a function of interaction
constant $\alpha _{s}$. The nearthreshold properties of regularized in a
mentioned way singular potential are examined. We obtain expressions for the
scattering lengths, which turn to be complex even for infinitesimal
imaginary part of interaction constant. The problem of perturbation of
nearthreshold states of regular potential by a singular one is treated, the
expressions for level shifts and widths are obtained. The physical sense of
suggested regularization is that the scattering observables are insensitive
to any details of the short range modification of singular potential, if
there exists sufficiently strong inelastic short range interaction. In this
case the scattering observables are determined by the solutions of
Schrodinger equation with regularized potential $-(\alpha _{s}\pm i0)/r^{s}$%
. Possible application of developed formalism to systems with short range
annihilation are discussed.
\end{abstract}


\section{Introduction.}

Singular potentials and related collapse problem is of special interest in
many fields. Attractive singular potentials are concerned in the three-body
problem in connection with famous Efimov and Thomas effect \cite{efimov}, in
nuclear physics (low energy $NN$ and $N\bar{N}$ interaction \cite{Phil,CPS}
), relativistic equations (Dirac equation with strong Coulomb potential and
more singular potentials \cite{popov}), atomic and molecular physics (van
der Waals interaction \cite{atomic}) and other fields. The review of
results, devoted to this one of the oldest quantum mechanical problem can be
found in \cite{SingPotRev1,SingPotRev2}.

The collapse or ''fall to the center'' \cite{LL}, produced by singular
potential, means that the energy spectrum is not bounded from below. The
wave function has infinite number of zeroes in the vicinity of the origin,
while two independent solutions of Schrodinger equation have the same order
of singularity and there is no obvious way to choose their linear
combination and define scattering observables.

The common way to deal with the pathological behavior of singular potentials
is to impose that exact physical short range interaction is not that
singular and the knowledge of the short-range parameters of interaction
enables unique definition of scattering observables. Following this logic
numerous methods of singular potentials regularization have been suggested 
\cite{renorm1,renorm2}. The obtained results are usually very strongly
depend on regularizing parameters (such as cut-off radius).

Another approach to the problem of collapse, which we will follow in this
paper, is to exploit the relation between singular potentials and inelastic
scattering \cite{Abs1,Abs2}. The aim of the present study is to investigate
the physical content of such a relation. The link between singular
potentials and inelasticity arises from the fact that Hamiltonian with real
attractive singular potential is not self-adjoint \cite{Math}. To make this
link explicit, we allow interaction constant to be complex $\alpha
_{s}=\alpha _{s}\pm i\omega $ ($\omega >0$). First we show that there is no
collapse for complex values of interaction constant. This follows from the
simple observation, that one of the solutions of Schrodinger equation
sharply decreases, while the other solution increases near the origin inside
the domain where complex potential acts \cite{KPV}. Thus one can choose
between independent solutions and calculate observables. Less trivial is
that even in case there is an infinitesimal imaginary part of $\alpha
_{s}=\alpha _{s}\pm i0$, the collapse is eliminated and observables are
uniquely defined. The non-self-adjointness of Hamiltonian for real positive
values of interaction constant $\alpha _{s}$ manifests itself in the fact
that S-matrix is not unitary even for infinitesimal imaginary part of
interaction constant. The S-matrix module is less than unit for $\alpha
_{s}=\alpha _{s}+i0$ (absorption) and greater than unit for $\alpha
_{s}=\alpha _{s}-i0$ (creation). In this sense the scattering on a singular
potential can be \textquotedblright redefined\textquotedblright\ as a
scattering on an absorptive (creative) potential with interaction constant $%
\alpha _{s}=\alpha _{s}\pm i0$. We will show that the physical meaning of
such an approach is that the scattering observables can be uniquely defined
and are independent on any details of short range modification of singular
potential, if this short range interaction includes sufficiently strong
inelasticity. We mention possible application of developed formalism to
systems with short range annihilation, including nucleon-antinucleon and
atom-antiatom systems.

\subsection{Potentials more singular than $-\protect\alpha /r^{2}$.}

In the following we put $2M=1$. Essential point is that interaction constant
is considered to be complex $\alpha _{s}=\alpha _{s}\pm i\omega $. Near the
origin we can neglect energy and all the nonsingular potentials, increasing
at the origin slower than $1/r^{2}$, so the Schrodinger equation becomes:

\[
\left[ -\frac{\partial ^{2}}{\partial r^{2}}+\frac{l(l+1)-\alpha _{2}}{r^{2}}%
-\frac{\alpha _{s}}{r^{s}}\right] \Phi (r)=0 
\]

The following variable substitution in the wave-function $\Phi (r)$:

\begin{eqnarray}
\Phi (r) &=&\sqrt{r}K(z)  \label{Vsubst} \\
z &=&\frac{2\sqrt{\alpha _{s}}}{s-2}\text{ }r^{-(s-2)/2}\text{ }
\end{eqnarray}
yields in Bessel equation:

\begin{eqnarray}
K^{\prime \prime }+\frac{1}{z}K^{\prime }+(1-\frac{\mu ^{2}}{z^{2}})K &=&0 
\nonumber \\
\mu &=&\frac{\sqrt{(2l+1)^{2}-4\alpha _{2}}}{s-2}
\end{eqnarray}

It's solution is: 
\begin{equation}
K=CH_{\mu }^{(1)}(z)+DH_{\mu }^{(2)}(z)
\end{equation}

Here $H_{\mu }^{(1)}(z)$ and $H_{\mu }^{(2)}(z)$ are Hankel functions of
order $\mu $ \cite{Watson}. It is worth to mention that the new variable $%
z=2p_{eff}r/(s-2)$ is a semiclassical phase, and local momentum $p_{eff}=%
\sqrt{\alpha _{s}/r^{s}+\alpha _{2}/r^{2}-l(l+1)/r^{2}}$ coincides with
classical momentum $p_{eff}\approx \sqrt{\alpha _{s}}/r^{s/2}$ near origin.

Let us cut the singular potential at small distance $r_{0\text{ }}$, and
replace it with square well: 
\begin{equation}
U(r)=\left\{ 
\begin{array}{cll}
-\alpha _{s}/r^{s} & \mbox{if} & r>r_{0} \\ 
-\alpha _{s}/r_{0}^{s} & \mbox{if} & r\leq r_{0}%
\end{array}
\right.  \label{cut}
\end{equation}

The logarithmic derivative of the wave-function in the square well at point $%
r_{0}$ is:

\begin{equation}
\frac{\Phi ^{\prime }}{\Phi }\rfloor _{r=r_{0}}=p_{eff}\cot (p_{eff}r_{0})
\label{LD1}
\end{equation}

We can choose $r_{0}$ small enough to ensure:

\[
\left| p_{eff}r_{0}\right| \approx \sqrt{\left| \alpha _{s}\right| }%
/r^{(s-2)/2}\gg 1 
\]
As far as $\alpha _{s}$ is complex (for distinctness we fix $\mathop{\rm Im}%
\alpha _{s}>0$) we have: 
\[
\mathop{\rm Im}%
p_{eff}r_{0}\gg 1 
\]

Then, taking into account that in this case $\cot (p_{eff}r_{0})\rightarrow
-i$, we get for the mentioned above logarithmic derivative

\[
\frac{\Phi ^{\prime }}{\Phi }\rfloor _{r=r_{0}}=-ip_{eff} 
\]

which corresponds to the condition of full absorption of particle with
momentum $p_{eff}$ at distance $r_{0}$.

The logarithmic derivative of the wave-function in singular potential for $%
r>r_{0}$ is: 
\[
\frac{\Phi ^{\prime }}{\Phi }=-ip_{eff}\frac{\exp (iz-i{\pi /4-i\mu \pi /2)-%
\widetilde{S}\exp }(-iz-i{\pi /4-i\mu \pi /2)}}{\exp (iz-i{\pi /4-i\mu \pi
/2)+\widetilde{S}\exp }(-iz-i{\pi /4-i\mu \pi /2)}} 
\]
We used here the asymptotic expansion of Hankel functions of large argument.
Matching logarithmic derivatives we get for $\widetilde{S}\equiv D/C$: 
\[
\widetilde{S}=0 
\]

Thus the solution near the origin is: 
\begin{equation}
\Phi =\sqrt{r}H_{\mu }^{(1)}(\frac{2\sqrt{\alpha _{s}}}{s-2}r^{-(s-2)/2})
\label{Sol}
\end{equation}
which means that only ''incoming'' wave presents in the solution near origin
(it is convenient to define ''incoming'' wave as a solution of Schrodinger
equation, which logarithmic derivative at point $r$ is equal to $%
-ip_{eff}(r) $). For $\omega >0$ this wave-function rapidly decreases with $%
r\rightarrow 0 $.

The value of $\widetilde{S}$ uniquely defines all the scattering properties
of our potential. In particular we can now obtain S-wave scattering length
for potential $-\alpha _{s}/r^{s}$ (for $s>3$) from the asymptotic of the
wave-function at large $r$ (small $z$):

\begin{eqnarray}
H_{\mu }^{(1,2)} &=&\mp i\left[ (z/2)^{-\mu }/\Gamma (1-\mu )-\exp (\mp i\pi
\mu )\text{ }(z/2)^{\mu }/\Gamma (1+\mu )\right] /\sin (\mu \pi ) \\
\Phi &\sim &1-r/a_{0}
\end{eqnarray}

\begin{equation}
a_{S}^{abs}=\exp (-i\pi /(s-2))\left( \frac{\sqrt{\alpha _{s}}}{s-2}\right)
^{2/(s-2)}\Gamma ((s-3)/(s-2))/\Gamma ((s-1)/(s-2))  \label{sclength}
\end{equation}

The above results are valid in the limit $r_{0}\rightarrow 0$.

We can now go to the limit $\omega \rightarrow +0$, as far as $\widetilde{S}$
is independent on $\omega $.We must ensure that $%
\mathop{\rm Im}%
z=2/(s-2)%
\mathop{\rm Im}%
pr_{0}\gg 1$ which is achieved by the order of performing limiting
transitions. First we calculate $\tilde{S}$ for nonzero $\omega $ (and put $%
r_{0}\rightarrow 0$), and then put $\omega \rightarrow +0$. Let us note,
that according to our cut-off procedure (\ref{cut}) the singular potential
is real everywhere except infinitesimal vicinity of the origin.

It is interesting, that in spite $%
\mathop{\rm Im}%
\alpha \rightarrow +0$ the scattering length has nonzero imaginary part. As
we've mentioned above, this is the manifestation of the singular properties
of potential which violates the self-adjointness of Hamiltonian.

For $\omega =%
\mathop{\rm Im}%
\alpha _{s}\rightarrow -0$ one can easily get $\widetilde{S}=\infty $ and
mentioned S-wave scattering length for creation is:

\[
a_{S}^{cr}=\left( a_{S}^{abs}\right) ^{\ast }=\exp (i\pi /(s-2))\left( \frac{%
\sqrt{\alpha _{s}}}{s-2}\right) ^{2/(s-2)}\Gamma ((s-3)/(s-2))/\Gamma
((s-1)/(s-2)) 
\]

Thus the scattering length has well-determined and conjugated values in the
upper and lower complex halfplanes of $\alpha _{s}$.

If we turn to (\ref{sclength}), we see, that the real part of S-wave
scattering length is a fixed sign function for any positive value of $\alpha
_{s}$, which indicates there are no bound states in regularized singular
potential $-(\alpha _{s}\pm i0)/r^{s}$ (we return to the spectrum problem of
singular potential further on). Let us compare the scattering length (\ref%
{sclength}) with that of the repulsive singular potential $\alpha _{s}/r^{s}$%
. One can get: 
\begin{equation}
a_{S}^{rep}=\left( \frac{\sqrt{\alpha _{s}}}{s-2}\right) ^{2/(s-2)}\Gamma
((s-3)/(s-2))/\Gamma ((s-1)/(s-2))  \label{screpulsive}
\end{equation}

It is easy to see, that (\ref{sclength}) can be obtained from (\ref%
{screpulsive}) simply by choosing certain branch (corresponding to
absorption or creation) of the function $\left( \sqrt{\alpha _{s}}\right)
^{2/(s-2)}$ when passing through the branching point $\alpha _{s}=0$. The
scattering length in singular potential is an analytical function of $\alpha
_{s}$ in the whole complex \ plane of $\alpha _{s}$ with a cut along
positive real axis. The jump of scattering length is:

\[
\Delta a\equiv a_{S}^{abs}-a_{S}^{cr}=-2i\sin (\frac{\pi }{s-2})\left( \frac{%
\sqrt{\alpha _{s}}}{s-2}\right) ^{2/(s-2)}\Gamma ((s-3)/(s-2))/\Gamma
((s-1)/(s-2)) 
\]

It is also curious that the real part of regularized scattering length for $%
3<s<4$ is negative, which corresponds to attraction, while for $s>4$ it
becomes positive, which corresponds to repulsion. One can say that strong
absorption (creation) in the origin results in effective repulsion. The
attractive effect of the potential tail dominates over repulsive effect of
the origin only for $s<4$. For more rapidly decreasing potentials with $s>4$
the overall effect is repulsive. In case of $s=4$ there is exact
compensation of core repulsion and tail attraction, so $%
\mathop{\rm Re}%
a=0$.

It is worth to mention, that above results can be obtained by fixing $%
\alpha_s $ real, but making infinitesimal complex rotation of $r$-axis:

\[
r=\rho \exp (i\varphi ) 
\]

which yields in Schrodinger equation:

\[
\left[ -\frac{\partial ^{2}}{\partial \rho ^{2}}+\frac{l(l+1)-\alpha _{2}}{%
\rho ^{2}}-\frac{\alpha _{s}(1-i(s-2)\varphi )}{\rho ^{s}}\right] \Phi (\rho
\exp (i\varphi ))=0 
\]

One can see that negative values of $\varphi $ correspond to absorption,
while positive values correspond to creation.

Summarizing, we can say that the scattering on attractive singular potential 
$-\alpha _{s}/r^{-s}(s>2)$ can be defined as full absorption (creation) of
the particle in such a potential in the origin. This procedure is supported
by the existence of the limit for $\widetilde{S}\equiv D/C$ (which defines
the linear combination of independent solutions) when interaction constant
is approaching real axis from upper (lower) complex halfplane: $\alpha
_{s}=\alpha _{s}\pm i0.$ The physical sense of such a regularization is as
follows. In case, when a singular interaction includes strong short range
inelastic component (the restrictions on the strength of such a short range
inelastic component would be discussed further on) the scattering
observables are insensitive to any details of the short range physics and
are determined by the solution (\ref{Sol}) of Schrodinger equation with
regularized singular potential $-(\alpha _{s}\pm i0)/r^{s}$.

\subsection{Potential $-\protect\alpha /r^{2}$}

Let us now turn to the very important case $-\alpha /r^{2}$. The
wave-function now is:

\begin{eqnarray}
\Phi &=&\sqrt{r}\left[ CJ_{\nu _{+}}(kr)+DJ_{\nu _{-}}(kr)\right] \\
\nu _{\pm } &=&\pm \sqrt{1/4-\alpha }
\end{eqnarray}

where $k=\sqrt{E}$, and $J_{\nu _{\pm }}$ are the Bessel functions\cite%
{Watson}. We use the same cut-off procedure (\ref{cut}) and put $\alpha
=\alpha +i\omega $.

Using Bessel function behavior at small $r$ we get:

\[
\Phi \sim r^{\nu _{+}+1/2}+\widetilde{S}r^{\nu _{-}+1/2} 
\]

where $\widetilde{S}\equiv D/C$. In the following we will be interested in
the values of $\alpha $ greater than critical $%
\mathop{\rm Re}%
\alpha >1/4.$ We mention that for $\nu _{+}=\sqrt{1/4-\alpha }$, $%
\mathop{\rm Re}%
\nu >0,%
\mathop{\rm Im}%
\nu <0$ , for small $\omega $ we find $%
\mathop{\rm Re}%
\nu =\omega /2\sqrt{%
\mathop{\rm Re}%
\alpha -1/4}$

Matching of logarithmic derivatives at cut-off point $r_0$ gives for $%
\widetilde{S}$:

\[
\widetilde{S}=r_{0}^{\nu _{+}-\nu _{-}}const\sim r_{0}^{2\nu }=r_{0}^{\omega
/\sqrt{%
\mathop{\rm Re}%
\alpha -1/4}}r_{0}^{-2i\sqrt{%
\mathop{\rm Re}%
\alpha -1/4}} 
\]

One can see, that due to the presence of the imaginary part of interaction
constant $\omega >0$ $\widetilde{S}\rightarrow 0$ when $r_{0}\rightarrow 0$.

Practically, this limit is achieved very slowly, i.e. if we want $\left| 
\widetilde{S}\right| <\varepsilon$, one needs to get $r_{0}<\varepsilon ^{%
\sqrt{%
\mathop{\rm Re}%
\alpha -1/4}/\omega }$

Thus, for $\alpha =\alpha +i0$ the wave-function is 
\begin{eqnarray}
\Phi &=&C\sqrt{r}J_{\nu _{+}}(kr) \\
\nu _{_{+}} &=&+\sqrt{1/4-\alpha }
\end{eqnarray}

For large argument this function behaves like: 
\[
\Phi \sim \cos (z-\nu _{_{+}}\pi /2-\pi /4) 
\]

The corresponding scattering phase is:

\begin{equation}
\delta =\frac{i\pi }{2}\sqrt{\alpha -1/4}-\pi /4  \label{scphase}
\end{equation}

One can see that the regularized wave-function and phase-shift are
analytical functions of $\alpha $ in the whole complex plane with a cut
along real axis $\alpha >1/4.$The jump of the phase-shift on the cut is:

\[
\Delta \delta =i\pi \sqrt{\alpha -1/4} 
\]

The mentioned procedure enables to choose between the independent solutions
of Schrodinger equation. In case of absorption ($\alpha =\alpha +i0)$ we
must choose the solution with negative imaginary index $\nu _{+}$, in case
of creation- with positive imaginary index $\nu _{-}$. We see, that
scattering on potential $-\alpha /r^{2}$ with $%
\mathop{\rm Re}%
\alpha >1/4,\omega \rightarrow \pm i0$ results in partial absorption
(creation), characterized by imaginary scattering phase (\ref{scphase}).

\subsection{Regularization of real singular potential by complex less
singular potential.}

As it was mentioned above, the essence of suggested regularization is that
the presence of strong enough inelastic interaction at short distance
suppresses one of the independent solutions of Schrodinger equation and
singles out another one. Now we would like to determine the minimum order of
singularity of infinitesimal imaginary potential required for regularization
of given singular potential. The potential of interest is a sum $-\alpha
_{s_{1}}/r^{s_{1}}-\alpha _{s_{2}}/r^{s_{2}}$, $s_{1}=s_{2}+t$. Here we keep 
$\alpha _{s_{1}}$ real, but put $\omega =$ $%
\mathop{\rm Im}%
\alpha _{s_{2}}>0.$

We again use the cut-off at some small $r_{0}$ and replace potential with
square well. The logarithmic derivative of square well wave-function for
small enough $r_{0}$ is:

\begin{eqnarray*}
\frac{\Phi ^{\prime }}{\Phi }\rfloor _{r=r_{0}} &=&p\cot (pr_{0}) \\
pr_{0} &=&\frac{\sqrt{\alpha _{s_{1}}}}{r^{(s_{1}/2-1)}}+\frac{i\omega \sqrt{%
1/\alpha _{s_{1}}}}{2r^{(s_{1}/2-1-t)}}
\end{eqnarray*}

\bigskip

One can see, that if $t<s_{1}/2-1$ than for $r_{0}\rightarrow 0$ $%
\mathop{\rm Im}%
pr_{0}\rightarrow +\infty $ and $p\cot (pr_{0})\rightarrow -ip$. Thus we
return to previously examined case, where we found that $S\equiv D/C=0.$ It
means, that we can put $r_{0}=0$ and than go to limit $\omega \rightarrow +0$
(in numerical calculations keeping $\omega \sqrt{1/\alpha _{s_{1}}}%
/2r^{(s_{1}/2-1-t)}\gg 1$).

If $t=s_{1}/2-1$ than for $r_{0}\rightarrow 0$ $%
\mathop{\rm Im}%
pr_{0}\rightarrow i\omega \sqrt{1/\alpha _{s_{1}}}/2$ and $p\cot
(pr_{0})\rightarrow p\cot (i\omega \sqrt{1/\alpha _{s_{1}}}/2)$

Using asymptotic behavior of Bessel functions, we get for $S$ :

\[
S=\frac{i\exp (i2pr_{0}/(s_{1}-2)-i{\pi /4-i\nu \pi /2)+}\cot (pr_{0})\exp
(i2pr_{0}/(s_{1}-2)-i{\pi /4-i\nu \pi /2)}}{\exp (i2pr_{0}/(s_{1}-2)-i{\pi
/4-i\nu \pi /2)\cot (pr_{0})-i\exp }(i2pr_{0}/(s_{1}-2)-i{\pi /4-i\nu \pi /2)%
}} 
\]

The value of $S$ oscillates with decreasing $r_{0}$ and there is no limit
for $r_{0}\rightarrow 0$, until $%
\mathop{\rm Im}%
pr_{0}\gg 1$ for which case we return to $S=0.$ So for $t=s_{1}/2-1$ the
regularization is possible only for large, noninfinitesimal $\omega \gg 
\sqrt{\alpha _{1}}$.

Obviously, for $t>s_{1}/2-1$ mentioned above regularization is impossible
since $%
\mathop{\rm Im}%
pr_{0}\rightarrow 0$ with $r_{0}\rightarrow 0.$

Summarizing the above results, we may say that the scattering is insensitive
to any details of regularizing short range interaction in singular potential 
$-\alpha _{s_{1}}/r^{s_{1\text{ }}}$if the inelastic component of such an
interaction behaves more singular than $-1/r^{(s_{1}/2+1)}$.

\subsection{Boundary condition}

As it follows from (\ref{Sol}) the regularization of singular potential by
means of infinitesimal imaginary addition to interaction constant is equal
to the boundary condition at the origin. For $s>2$:

\begin{eqnarray}
\Phi &\sim &\sqrt{r}H_{\mu }^{(1)}(\frac{2\sqrt{\alpha _{s}}}{s-2}%
r^{-(s-2)/2})  \label{BC} \\
\mu &=&\frac{\sqrt{(2l+1)-4\alpha _{2}}}{s-2}
\end{eqnarray}

\bigskip or:

\begin{equation}
\frac{\Phi ^{\prime }}{\Phi }\rfloor _{r\rightarrow 0}=-ip_{eff}  \label{BC1}
\end{equation}

\bigskip As one can see, this is the ''full absorption'' boundary condition.

For $s=2$: 
\begin{eqnarray}
\Phi &\sim &\sqrt{r}J_{\nu _{+}}(kr)  \label{BC2} \\
\nu _{\pm } &=&\pm \sqrt{1/4-\alpha }
\end{eqnarray}

or:

\begin{equation}
\frac{\Phi ^{\prime }}{\Phi }\rfloor _{r\rightarrow 0}=\frac{(1/2\mp i\left|
\nu _{\pm }\right| )}{\sqrt{1/4-\nu _{\pm }^{2}}}p_{eff}  \label{BC2p}
\end{equation}

This boundary condition corresponds to partial absorption (creation).

\subsection{\protect\bigskip Singular potential and WKB approximation.}

WKB approximation, consistent with the boundary condition (\ref{BC}) for $%
s>2 $ is : 
\begin{equation}
\Phi =\frac{1}{\sqrt{p_{eff}}}\exp (i\int\limits_{r}^{a}p_{eff}dr)
\label{semicl}
\end{equation}
One can easily check that the above expression coincides with an asymptotic
form of solution (\ref{Sol}) for small $r$ (large $z$).

The WKB approximation holds if:

\[
|\frac{\partial (1/p_{eff})}{\partial r}|\ll 1 
\]

In case of zero-energy scattering on singular potential with $s>2$ this
condition holds for:

\[
r\ll (2\sqrt{\alpha _{s}}/s)^{(2/s-2)} 
\]

i.e. near origin.

For $s=2$ the semiclassical approximation is valid only for $\alpha \gg 1$.
One can see that in this case the boundary condition (\ref{BC2p}) becomes
the condition of full absorption (creation) (\ref{BC1}).

We can conclude, (for distinctness we will speak here of absorptive
potentials with $\alpha =\alpha +i0$) in case WKB approximation is valid
everywhere the Schrodinger equation solution includes incoming wave only.
The corresponding S-matrix $S=0$ within such an approximation.

On the other hand nonzero value of S-matrix means that an outgoing wave
presents in the solution. The outgoing wave can appear in the solution only
in the regions where (\ref{semicl}) does not hold. For example, in the zero
energy limit $E\rightarrow 0$ the S-matrix is nonzero: 
\begin{equation}
S=\frac{1-ika_{S}^{abs}}{1+ika_{S}^{abs}}
\end{equation}
with $a_{S}^{abs}$ from (\ref{sclength}) and $k=\sqrt{E}$. One can show that
the outgoing wave is reflected from those parts of potential which change
sufficiently fast in comparison with effective wavelength: 
\[
|\frac{\partial (1/p_{eff})}{\partial r}|\gtrsim 1 
\]
For zero energy scattering and $l=0$ this holds for

\[
r\geq (2\sqrt{\alpha _{s}}/s)^{(2/s-2)} 
\]

\subsection{Remarks on the spectrum of singular potential.}

\subsubsection{\protect\bigskip Spectrum of potential more singular than $-%
\protect\alpha _{2}/r^{2}$.}

The ''infinitely deep'' bound states produced by singular potential is the
central point of the collapse problem.

A simple qualitative picture of the energy spectrum of singular potential is
as follows. One can have an estimation $E_{apr}$ from above for the ground
state $E_{0}$ in a singular potential , replacing singular potential by a
square well at small $r_{0\text{ }}$. Such an estimation is a ground state
energy in the mentioned square well with depth $\alpha _{s}/r_{0}^{s}$ and
width $r_{0}$ . For small enough $r_{0}$:

\begin{eqnarray}
E_{apr} &\simeq &-\alpha _{s}/r_{0}^{s}+\pi ^{2}/r_{0}^{2} \\
E_{apr} &\rightarrow &-\infty \text{ }r_{0}\rightarrow 0
\end{eqnarray}

This ensures that exact level $E_{0} < E_{apr}$ also tends to minus infinity
with decreasing of $r_{0}$.

Our regularizing procedure $\alpha _{s}\rightarrow \alpha _{0}\pm i\omega $
results in:

\[
E_{apr}\simeq -\alpha _{0}/r_{0}^{s}\pm i\omega /r_{0}^{s}+\pi
^{2}/r_{0}^{2} 
\]

As far as we always keep $%
\mathop{\rm Im}%
p_{eff}r_{0}=\pm \omega r_{0}^{-(s-2)/2}\sqrt{1/\alpha _{s}}\rightarrow
\infty $ the width of the \textquotedblright ground\textquotedblright\ state
would be infinitely large even in case $\omega \rightarrow \pm 0$.

This infinitely large width corresponds to the full absorption (creation) of
the particle in the origin. One can say that the collapse is eliminated
because the particle disappears before it approaches close enough to the
scattering center. Practically this means, that there are no bound states in
the regularized in a mentioned above way singular potential. Absence of
bound states is clear also from the already mentioned fact, that scattering
amplitudes in regularized attractive singular potential can be obtained by
an analytical continuation in $\alpha _{s}$ of corresponding scattering
amplitude in repulsive singular potential.

\subsubsection{\protect\bigskip Spectrum of potential $-\protect\alpha %
_{2}/r^{2}.$}

The S-matrix in potential $-(\alpha _{2}\pm i0)/r^{2}$ , according to (\ref%
{scphase}) is energy independent:

\[
|S|=\exp (\mp \pi \sqrt{\alpha _{2}-1/4}) 
\]

Obviously, such an S-matrix has no singularities and there are no bound
states in such a potential.

Meanwhile, if $-(\alpha _{2}\pm i0)/r^{2}$ is combined with another
potential, which can produce bound states itself, these bound states may
obtain finite width. Let us treat, for example, the case of attractive
coulomb potential, which is combined with potential $-(\alpha _{2}+i0)/r^{2}$

It is easy to see, that the energy is:

\begin{eqnarray*}
E &=&-\frac{1}{2n^{2}} \\
n &=&n_{r}+\nu _{+}+1/2
\end{eqnarray*}

where radial quantum number $n_{r}$ is positive integer or zero.

Finally, for $\alpha >1/4$:

\[
E=-\frac{1}{2}\frac{n_{r}+1/2}{n_{r}^{2}+n_{r}+\alpha }-\frac{i}{2}\frac{%
\sqrt{\alpha -1/4}}{n_{r}^{2}+n_{r}+\alpha } 
\]

Further we will examine in more details the modification of spectrum of the
regular potential by a singular one.

\subsection{Nearthreshold scattering and perturbation theory}

We will be interested in the modification of the low energy scattering
amplitude of regular potential $U(r)$, by a potential, which has singular
behavior $-(\alpha _{s}+i0)/r^{s}-\alpha _{2}/r^{2}$ at short distances. We
will also treat the modification of the spectrum of nearthreshold states in
such a potential. If $\alpha _{s}$ is small enough, there is a range where 
\begin{equation}
U(r)\ll \alpha _{s}/r^{s}\ll (l(l+1)-\alpha _{2})/r^{2}  \label{domain}
\end{equation}%
Let us first treat the case, when regular potential is approximately
constant in the mentioned range $U(r)\approx p^{2}$. Then from (\ref{domain}%
) we get:

\begin{equation}
p\alpha _{s}^{1/(s-2)}\ll 1  \label{pro}
\end{equation}

For such values of $r$ the wave function is:

\begin{eqnarray}
&&\Phi \sim \sqrt{r}(J_{\mu }(pr)-\tan (\delta _{s})Y_{\mu }(pr))
\label{Free} \\
Y_{\mu } &=&\frac{J_{\mu }\cos (\mu \pi )-J_{-\mu }}{\sin (\mu \pi )} \\
&&\mu =\sqrt{(l+1/2)^{2}-\alpha _{2}}  \nonumber
\end{eqnarray}

here $\delta _{s}$ is a phase shift, produced by singular potential in the
presence of regular potential.

For small $r\precsim \alpha _{s}^{1/(s-2)}$ the wave-function is determined
by singular and centrifugal potential:

\begin{eqnarray*}
\Phi &\sim &\sqrt{r}H_{\nu }^{+}(\frac{2\sqrt{\alpha }}{s-2}r^{-(s-2)/2}) \\
\nu &=&2\mu /(s-2)
\end{eqnarray*}

Matching logarithmic derivatives and taking into account (\ref{pro}) we get
for the phase shift $\delta _{s}$:

\[
\delta _{s}=-\sin (\pi \mu )(\frac{p\alpha _{s}^{1/(s-2)}}{2(s-2)^{2/(s-2)}}%
)^{2\mu }\exp (-i\pi \nu )\frac{\Gamma (1-\mu )\Gamma (1-\nu )}{\Gamma
(1+\mu )\Gamma (1+\nu )} 
\]

which for integer values of $2\mu =2l+1$ becomes:

\[
\delta _{s}=(-1)^{l+1}(\frac{p\alpha _{s}^{1/(s-2)}}{2(s-2)^{2/(s-2)}}%
)^{2l+1}\exp (-i\pi \nu )\frac{\Gamma (1/2-l)\Gamma (1-\nu )}{\Gamma
(3/2+l)\Gamma (1+\nu )} 
\]

Let us mention, that for nonzero $l$ the value of $%
\mathop{\rm Re}%
\delta _{s}$ may become smaller, than the correction to expression (\ref%
{Free}), produced by the tail of singular potential, which is now small in
comparison with $U(r)$. Such a correction depends on certain form of the
tail of singular potential and can be calculated as a first order of
distorted wave approximation.

In the same time $\delta _{s}$ has positive imaginary part according to
absorbing character of singular potential.

\begin{equation}
\mathop{\rm Im}%
\delta _{s}=(-1)^{l}(\frac{p\alpha _{s}^{1/(s-2)}}{2(s-2)^{2/(s-2)}}%
)^{2l+1}\sin (\pi \nu )\frac{\Gamma (1/2-l)\Gamma (1-\nu )}{\Gamma
(3/2+l)\Gamma (1+\nu )}  \label{Imdelta}
\end{equation}

In case of coulomb potential $U(r)=-\beta /r$, the solution in singular
potential should be matched with zero-energy coulomb wave-function:

\begin{eqnarray*}
&&\Phi _{c} \sim \sqrt{r}(J_{\eta }(\sqrt{8\beta r})-\tan (\delta
_{s})Y_{\eta }(\sqrt{8\beta r})) \\
&&\eta =2\sqrt{(l+1/2)^{2}-\alpha _{2}}
\end{eqnarray*}

The expression for the phase shift now is:

\[
\delta _{s}=-\sin (\pi \eta )(\frac{8\beta \alpha _{s}^{1/(s-2)}}{%
2(s-2)^{2/(s-2)}})^{\eta }\exp (-i\pi \eta )\frac{\Gamma ^{2}(1-\eta )}{%
\Gamma ^{2}(1+\eta )} 
\]

In the upper expression $\eta $ is noninteger. For the validity of the above
expression it is required that $\beta \alpha _{s}^{1/(s-2)}\ll 1$.

From the above results we can immediately get the low energy phase shift and
scattering volume $a_{l}$ produced by pure singular potential $-(\alpha
_{s}+i0)/r^{s}$, putting $U(r)=k^{2}=E$.

\begin{eqnarray*}
\delta _{s} &=&(-1)^{l+1}(\frac{k\alpha _{s}^{1/(s-2)}}{2(s-2)^{2/(s-2)}}%
)^{2l+1}\exp (-i\pi \nu )\frac{\Gamma (1/2-l)\Gamma (1-\nu )}{\Gamma
(3/2+l)\Gamma (1+\nu )} \\
a_{l} &=&(-1)^{l}(\frac{\alpha _{s}^{1/(s-2)}}{2(s-2)^{2/(s-2)}})^{2l+1}\exp
(-i\pi \nu )\frac{\Gamma (1/2-l)\Gamma (1-\nu )}{\Gamma (3/2+l)\Gamma (1+\nu
)}
\end{eqnarray*}

The nearthreshold states produced by regular potential $U(r)$ are perturbed
by the short range singular potential. In particular they get the widths,
which in our case of small $\delta _{s}$ are proportional to $%
\mathop{\rm Im}%
\delta _{s}$.

If the nearthreshold states spectrum in $U(r)$ has semiclassical character,
than from the quantization rule:

\[
\int \sqrt{E_{n}+\delta E_{n}-U\left( r\right) }dr+\delta _{s}=const 
\]

one gets:

\begin{equation}
\delta E_{n}=-\delta _{s}\omega _{n}  \nonumber
\end{equation}

where $\omega _{n}$ is semiclassical frequency:

\[
\omega _{n}=(\int (E_{n}-U\left( r\right) )^{-1/2}dr)^{-1} 
\]

Taking into account (\ref{Imdelta}) we get for the width of the state:

\[
\Gamma _{n}/2=(-1)^{l+1}(\frac{p\alpha _{s}^{1/(s-2)}}{2(s-2)^{2/(s-2)}}%
)^{2l+1}\sin (\pi \nu )\frac{\Gamma (1/2-l)\Gamma (1-\nu )}{\Gamma
(3/2+l)\Gamma (1+\nu )}\omega _{n} 
\]

We see that singular potential modification of nearthreshold scattering in
regular potential is small when the characteristic wave-length of the
particle in the regular potential $1/p$ is much greater than scattering
length in singular potential, so that $p\alpha _{s}^{1/(s-2)}\ll 1$.

\subsection{Physical examples.}

\subsubsection{Nearthreshold nucleon-antinucleon scattering.}

The meson-exchange inspired models of nucleon-antinucleon low energy
interaction \cite{Phil,CPS,KW,DR,Pign} encounter attractive potentials that
behave like $1/r^{3}$ near origin. These models require short-range
regularization of singular attractive potentials. Such a regularization is
usually done by means of cut-off radius. Another important ingredient of
mentioned models is a short-range imaginary potential, which describes
nucleon-antinucleon annihilation. We will demonstrate the regularization of
attractive singular potentials by means of imaginary addition to the
strength parameter of singular potential. The short-range annihilation is
automatically taken into account in our approach due to mentioned above full
absorption of the particle in the vicinity of the scattering center. In the
same time no cut-off radius is needed. We calculate the scattering volume in
the state with quantum numbers $J=0$, $S=1$, $L=1$, $T=0$, where singular $%
1/r^{3}$ terms are attractive. It is known that this particular scattering
state has a nearthreshold resonance, so the scattering volume is very much
enhanced. The reproduction of such an enhancement can be a test for the
model. We use the version of real OBEP potential $W_{OBEP}$, used in
Kohno-Weise model \cite{KW}. This potential includes singular terms at short
distance, which are now regularized by introducing imaginary singular
potential of the form $W_{I}=-i\omega /r^{3}\exp (-r/\tau )$ with $\omega
\rightarrow 0$:

\[
W=W_{OBEP}-i\omega /r^{3}\exp (-r/\tau ) 
\]

The scattering volume $^{3}P_{0\text{ }}T=0$ calculated in the limit $\omega
\rightarrow 0$ turns to be:

\[
a_{r}=-7.66-i4.87\text{ fm}^{3} 
\]

The result is independent on diffuseness $\tau $, as far as $\omega $ is
infinitesimal.

The value of the same scattering volume, obtained within Kohno-Weise model
with a cut-off $r_{c}=1$fm is:

\[
a_{KW}=-8.83-i4.45\text{ fm}^{3} 
\]

As one can see, both scattering volumes are rather close, reproducing the
strong P-wave enhancement. In the same time suggested approach is free from
any uncertainty related to the cut-off radius. The above result demonstrates
that condition of full absorption, incorporated in our model, is rather
realistic in case of $N\overline{N}$ interaction \cite{DM}.

\subsubsection{Hydrogen-antihydrogen interaction.}

It is known, that long range interaction between atoms is dominated by
attractive van der Waals potential $-C_{6}/r^{6}$. The very simple model of
low energy (fraction of eV) hydrogen-antihydrogen interaction is the
absorption model, which is based on the assumption that the particles are
fully absorbed on the sphere of radius $r_{c}$, while at $r>r_{c}$ the atoms
interact via $-C_{6}/r^{6}$ potential. The critical radius is close to
hydrogen Bohr radius $r_{c}\approx r_{B}$.

The scattering observables in such a model can be obtained within suggested
regularization of attractive singular potential by imaginary addition to the
interaction constant $\widetilde{C}_{6}=C_{6}(1+i\omega ).$ In the limit $%
\omega \rightarrow 0$ (which corresponds to $r_{c}\rightarrow 0$) we get for
S-wave scattering length:

\[
a_{S}=\left( MC_{6}\right) ^{1/4}\frac{\Gamma (3/4)}{2\sqrt{2}\Gamma (5/4)}%
(1-i) 
\]

Here $M$ is a proton mass.

We would not discuss here the physics of hydrogen-antihydrogen interaction 
\cite{Froel,Vor} and limitation of absorption model. We only mention, that
strong absorption model seems to be more realistic for the interaction of $%
1S $ hydrogen with excited $nS$ antihydrogen, where transitions to a lot of
inelastic channels are energetically allowed even in zero energy limit. If
we take into account that $C_{6}\sim n^{4}$, we get a useful law:

\[
a_{S}\sim n 
\]

Let us summarize that in mentioned above physical examples the strong
absorption makes the system insensitive to any details of short range
physics. This radically simplifies the problem and makes useful the
developed approach of singular potentials regularization by means of
imaginary addition to interaction constant.

\subsection{Conclusion.}

We have found that the scattering amplitude in singular potential $-\alpha
_{s}/r^{s}(s\geq 2)$ has well-defined values if interaction constant is
complex $\alpha _{s}=\alpha _{s}\pm i\omega $ $(\omega >0)$. This is true
even in the limiting case $\omega \rightarrow +0$. The corresponding
S-matrix is nonunitary ($|S|<0$ for $\omega >0$ and $|S|>0$ for $\omega <0$%
), has conjugated values in the complex halfplanes of $\alpha _{s}$ and
makes a jump on the real positive axis of $\alpha _{s}$. This property is a
manifestation of non-self-adjointness of Hamiltonian with singular potential
and can be used for \textquotedblright redefinition\textquotedblright\ of
singular potential as absorptive (creative) potential by adding an
infinitesimal imaginary part to the interaction constant $\alpha _{s}=\alpha
_{s}\pm i0$. Such a procedure is equal to boundary condition of absorption
(creation) of the particle in the origin. This procedure eliminates the
collapse of the system, as far as the particle promptly disappears in the
scattering center, rather than forms infinitely deep bound states. There are
no bound states in such a regularized singular potential. The scattering
observables in such a potential can be obtained from those of repulsive
singular potential by analytical continuation and choosing certain branch in
passing through the branching point $\alpha _{s}=0$.

It was shown, that scattering length perturbation theory can be used when we
are interested how the nearthreshold spectrum and scattering phase of
regular potential is modified by a singular one. The perturbation parameter
is the ratio of the scattering length in singular potential to the effective
wavelength in the regular potential near origin. We have obtained an
expression for the widths and shifts of mentioned states.

The physical sense of suggested regularization is that the scattering
observables are insensitive to any details of short range modification of
singular potential, if there is sufficiently strong short range inelastic
interaction. In this case the scattering amplitude can be calculated by
solving Schrodinger equation with the regularized singular potential $%
-(\alpha _{s}\pm i0)/r^{s}$. The developed formalism can be useful for
examining of particle-antiparticle systems with singular interaction and
short range annihilation.

\end{document}